\documentclass[prl, twocolumn]{revtex4}
\usepackage{amsmath}
\usepackage{graphicx}

\begin{document}

\preprint{}
\title{Numerical Study of the Spin Hall Conductance in the Luttinger Model}
\author{W.Q. Chen$^{1}$, Z.Y. Weng$^{1}$, and D.N. Sheng$^{2}$}
\affiliation{$^{1}$Center for Advanced Study, Tsinghua University, Beijing 100084\\
$^{2}$Department of Physics and Astronomy, California State University,
Northridge, CA 91330}

\begin{abstract}
We present first numerical studies of the disorder effect on the recently
proposed intrinsic spin Hall conductance in a three dimensional (3D) lattice
Luttinger model. The results show that the spin Hall conductance remains
finite in a wide range of disorder strength, with large fluctuations. The
disorder-configuration-averaged spin Hall conductance monotonically
decreases with the increase of disorder strength and vanishes before the
Anderson localization takes place. The finite-size effect is also discussed.
\end{abstract}

\maketitle

A primary goal of spintronics is to make use of spin degree of freedom of
electrons in the future `electronic' devices \cite{spintronics1,
spintronics2}. The spin Hall effect (SHE) may be one of potentially
effective ways to manipulate the spin transport. An extrinsic SHE generated
by impurities with spin-orbit (SO) coupling has been previously proposed
\cite{jetp}. By scattering electrons of different spins into
different directions, a net spin current can be established in the
transverse direction, accompanying the charge current induced by an applied
electric field. But usually the resulting spin accumulation is very weak as
it crucially depends on the impurity concentration. Recently, a much
stronger SHE due to the intrinsic SO coupling in clean materials has been
proposed for both the 3D p-doped semiconductors described by the Luttinger
model \cite{MNZ}, and the two-dimensional (2D) electron gas described by the
Rashba model \cite{Sinova1}. Here it has been argued that the
`dissipationless' spin currents can be of several orders of magnitude larger
than in the case of the extrinsic SHE. A signature of spin polarization
observed recently in the 2D hole gas (2DHG) \cite{exp1} and 3D n-doped
semiconductors \cite{exp2} might be originated from the intrinsic SHE \cite%
{2dhg,  2dsa}.

However, the effect of disorder on the intrinsic SHE remains a highly
controversial issue so far. It has been argued \cite{vertex1,vertex2} that
the spin current in the 2D Rashba model should vanish, even in the weak
disorder limit, after considering the vertex corrections. On the other hand,
it is shown that the vertex correction vanishes for the Luttinger \cite%
{vertex_luttinger} and 2DHG \cite{2dhg} models such that the SHE is robust
in the latter models at least when disorders are weak. Clearly disorder
effect is nonperturbative on the spin Hall transport properties, and
numerical approaches are highly desirable in order to
illustrate the fate of spin Hall conductance (SHC) in disordered systems.
So far there have been a
series of numerical works dealing with the SHE in the presence of disorders
in the 2D Rashba model \cite{LB1, LB3,kubo1,kubo2}. These calculations
suggest that the SHC survives  finite length scales in
disordered systems with indications of its  vanishing in the
thermodynamic limit. To our knowledge no numerical work has been done in the
3D Luttinger model with regard to the disorder effect.

In this Letter, we present a first numerical calculation of the
SHC in the lattice Luttinger model with including an on-site
random potential based on the Kubo formula. We find that the SHC at weak
disorder is intrinsically fluctuating, similar to the quantum Hall state
around the critical point. The distribution of the SHC over different
disorder configurations shows a strong symmetric peak with the averaged SHC
located at the peak position. The averaged SHC remains finite in a wide
range of disorder strengths covering the main regime of the metallic phase
while the finite-size scaling analysis suggests that SHC can survive at
larger length scales.
The calculated SHC decreases monotonically with
increasing disorder strength, and disappears not far before the 3D Anderson
localization takes place.

We start with the tight-binding version of the 3D Luttinger Hamiltonian,
which can be derived from the continuum version \cite{MNZ} with using the
replacement $k_{\nu }\rightarrow \sin k_{\nu }$ and $k_{\nu }^{2}\rightarrow
2(1-\cos k_{\nu })$. After a discrete Fourier transformation, the resulting
Hamiltonian reads
\begin{eqnarray}
H &=&-\sum_{\langle ij\rangle }(c_{i}^{\dagger }c_{j}+h.c.)+V_{L}\sum_{i,\nu
}(c_{i}^{\dagger }S_{\nu }^{2}c_{i+\nu }+h.c.)  \notag \\
&&+\frac{V_{L}}{8}\sum_{i,\mu \neq \nu }c_{i}^{\dagger }\{S_{\mu },S_{\nu
}\}(c_{i+\mu +\nu }+c_{i-\mu -\nu }  \notag \\
&&-c_{i+\mu -\nu }-c_{i-\mu +\nu })+\sum_{i}\epsilon _{i}c_{i}^{\dagger
}c_{i}
\end{eqnarray}%
where the electron annihilation operator $c_{i}$ has four components
characterized by the `spin'\ index $S_{z}=\frac{3}{2},\frac{1}{2},-\frac{1}{2%
},-\frac{3}{2},$ respectively, and $i+\nu $ ($\nu =x,y,z$) denote the
nearest-neighbors of site $i$, and $i+\nu +\mu $, etc., for the next
nearest-neighbor sites. Here $V_{L}\equiv \frac{2\gamma _{2}}{\gamma _{1}+%
\frac{5}{2}\gamma _{2}}$ represents the strength of the Luttinger
spin-orbital coupling. We choose $V_{L}=0.364$ as the ratio $\gamma
_{1}/\gamma _{2}$ is around 3 in typical semiconductors \cite{haug}. The
last term accounts for on-site nonmangetic disorder with $\epsilon _{i}$
randomly distributed within $[-W/2,W/2]$. Note that the Luttinger model is
only a valid description of real semiconductors around the $\Gamma $ points
at $k_{\nu }\rightarrow 0,$ which corresponds to choosing the Fermi energy
near the band edge in the present tight-binding version.

The SHC for \emph{each} disorder configuration can be calculated by the Kubo
formula\cite{kubo0}
\begin{equation}
\sigma _{SH}^{1}=-\frac{2}{N_L}\mathrm{Im}\sum_{E_{n}<E_{f}<E_{m}}\frac{%
\langle n|j_{x}^{y\mathrm{spin}}|m\rangle \langle m|j_{z}|n\rangle }{%
(E_{m}-E_{n})^{2}},  \label{eq:kubo_for_one}
\end{equation}%
in which $N_L$ is the number of lattice sites, $E_{f}$ denotes the Fermi
energy, $E_{m,n}$ is the eigen-energy, the charge current operator $\mathbf{j%
}=e\mathbf{v}$ and the spin current $\mathbf{j}_{\mu }^{\nu \mathrm{spin}}=%
\frac{1}{2}\{v_{\mu },S_{\nu }\}$. Here the velocity operator $\mathbf{v}$
as the conjugate operator of the position operator $\mathbf{R}\equiv
\sum_{i\sigma }\mathbf{r}_{i}n_{i\sigma } $ ($n_{i\sigma }$ is the number
operator at site $i$ with spin index $\sigma )$, is defined by the standard
relation $\mathbf{v}=\frac{i}{\hbar }[H,\mathbf{R}]$. In the presence of
random disorder $W\neq 0$, the SHC is obtained by averaging $\sigma _{SH}^{1}
$ over all disorder configurations, \emph{i.e.,}
\begin{equation}
\sigma _{SH}=\langle \sigma _{SH}^{1}\rangle .  \label{eq:kubo}
\end{equation}%
\begin{figure}[t]
\resizebox{80mm}{!}{    \includegraphics{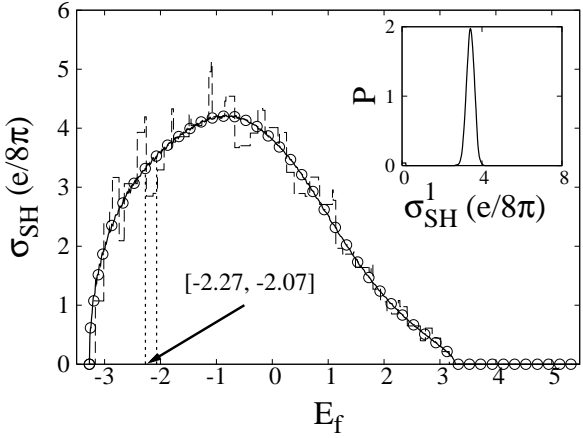} } \centering
\caption{$\protect\sigma _{SH}$ vs the Fermi energy $E_{f}$ in the pure
system. The dashed curve is for $8\times 8\times 8$ lattice and the solid
one is for $50\times 50\times 50$ lattice with PBC. The open circles for $%
8\times 8\times 8$ lattice are obtained by averaging over 200 configurations
of different BCs, which coincide with the solid curve very well. The inset
shows $P(\protect\sigma _{SH}^{1}),$ defined as the distribution of the spin
Hall conductance, at $8\times 8\times 8$ under different BCs with $E_{f}$'s
within the range indicated by the arrow in the main panel.}
\label{fig:pure_full_range}
\end{figure}
In a finite-size calculation, a proper boundary condition (BC) is
necessary for diagonalizing the Hamiltonian. A general (twisted)
BC, e.g., $\psi (x+L_{x},y,z) =e^{2\pi i\phi _{x}}\psi (x,y,z)$,
etc., where $L_{x }$ is the sample size along the $x$-direction,
and $\phi _{x}$ is defined within $[0,1]$ with $\phi _{x }=0$
corresponding to the periodic BC (PBC) along this direction. In
the thermodynamic limit, a physical quantity should not depend on
BCs. In a finite-size calculation, the BC averaging can be very
{\bf effective} in smoothening out finite-size fluctuations in
$\sigma _{SH}$ in a spin-orbit coupling system \cite{kubo2}. In
principle, this procedure is not necessarily the unique one for a
finite system (as one can also use the fixed BCs in the
calculation), but smoother data obtained this way can allow one to
make a finite-size scaling analysis and to meaningfully
extrapolate the results in the thermodynamic limit. For example,
let us first consider the PBC in the pure system with $\epsilon
_{i}=0$. The calculated $\sigma _{SH}$ for a $8\times 8\times 8$
lattice is shown in Fig. \ref{fig:pure_full_range} by the dashed
curve, which quickly fluctuates, as a function of $E_{f},$ with
finite steps due to the finite-size effect. Such a finite-size
effect disappears when the sample size is increased to $50\times
50\times 50$ (this size can only be reached for the pure system,
where the momentum is a good quantum number, in our calculation)
with the same PBC, as indicated by the smooth solid curve in Fig.
\ref{fig:pure_full_range}. On the other hand, if one averages over
different BCs (over 200 configurations) in Eq. \eqref{eq:kubo} for the $%
8\times 8\times 8$ lattice, the steps in the dashed curve can also become
smoothened out as represented by the open circles which coincide very well
with the solid curve obtained for the bigger lattice of $50\times 50\times
50 $ in Fig. \ref{fig:pure_full_range}.

The fluctuations in $\sigma _{SH}^{1}$ become very large in the presence of
disorders, typically in a range of 5-10 times larger than the averaged
value. To quantitatively describe such fluctuations, we shall introduce the
so-called distribution of the SHC (DSHC), $P(\sigma _{SH}^{1})$, which
determines the averaged SHC, $\sigma _{SH},$ by
\begin{equation}
\sigma _{SH}=\int d{\sigma _{SH}^{1}} P(\sigma _{SH}^{1})\sigma _{SH}^{1}.
\label{eq:p}
\end{equation}%
First, for a given $E_{f}$, we can calculate $\sigma _{SH}^{1}$ at different
disorder and BC configurations within a small Fermi energy interval, say, $%
[-2.27,-2.07]$ around $E_{f}=-2.17$ as illustrated in Fig. \ref%
{fig:pure_full_range} by the arrow [here\ the change in $\sigma _{SH}(E_{f})$
is presumably weak as a function of $E_{f}$]. Suppose that the total number
of computed $\sigma _{SH}^{1}$'s is $N$ in this range, and the number of $%
\sigma _{SH}^{1}$'s at $\sigma _{SH}^{1}=\sigma \pm \Delta \sigma $ [$\Delta
\sigma =\pm 0.01$ $(e/8\pi )]$, is denoted by $\Delta N(\sigma )$. Then the
DSHC is defined as the statistic distribution of $\sigma _{SH}^{1}$%
\begin{equation}
P(\sigma )=\frac{\Delta N(\sigma )}{N \Delta \sigma}.  \label{eq:dshc}
\end{equation}%
The DSHC for the pure system of a $8\times 8\times 8$ lattice for $200$
different BCs is shown in the inset of Fig. \ref{fig:pure_full_range}, in
which $P(\sigma )$ is a very symmetric peak such that one may simply use the
peak position, $\frac{3.4e}{8\pi }$, to determine the averaged $\sigma _{SH}$
instead of directly evaluating the average in Eq.(\ref{eq:p}). A similar
technique has been used in the quantum Hall effect system \cite{Sora_cho}.
\begin{figure}[t]
\resizebox{80mm}{!}{    \includegraphics{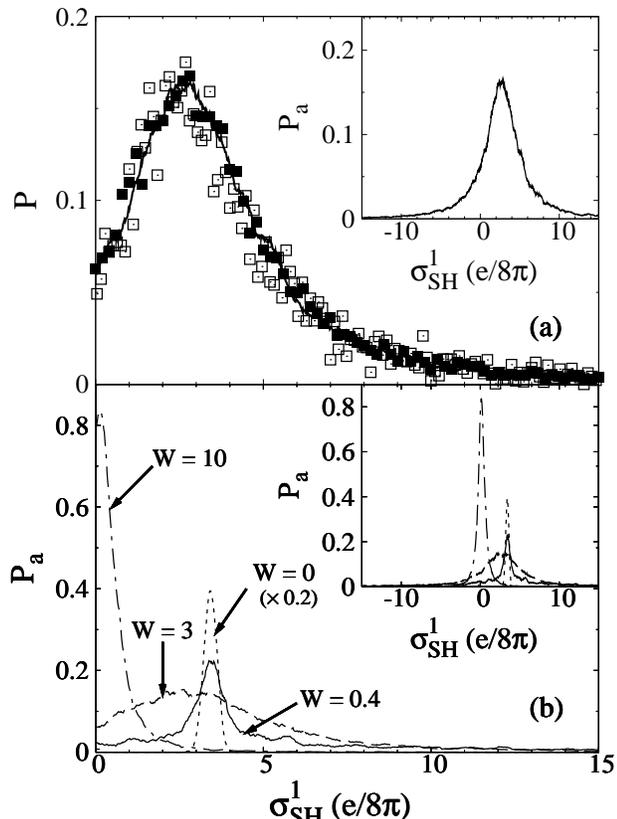} } \centering
\caption{(a) The DSHC at $W=3$ on a $6\times 6\times 6$ lattice. Open square
denotes 1000 random disorder and BC configurations, and the closed square is
for 5000 configurations. The solid curve is the averaged DSHC, $P_{a}$
(defined in the text), at 1000 configurations. The inset shows $P_{a}$ in a
larger scale. (b): $P_{a}$ at different disorder strengths. The dotted one
is for $W=0$; the solid curve: $W=0.4$; the dashed curve: $W=3;$ the
dash-dot curve: $W=10$. Inset: the same curves in a larger scale.}
\label{fig:various_impurity}
\end{figure}

\begin{figure}[t]
\resizebox{60mm}{!}{    \includegraphics{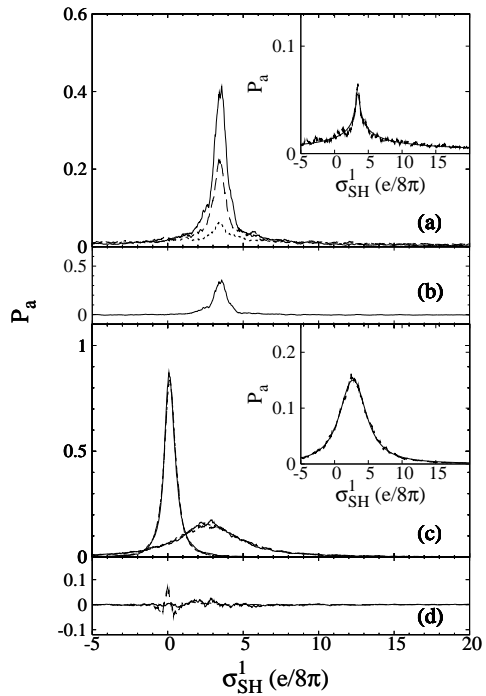} } \centering
\caption{(a). The DSHC at $W=0.4.$ The solid curve: a $6\times 6\times 6 $
lattice with 500 configurations, the dashed curve: a $8\times 8\times 8$
lattice with 200 configurations and the dotted curve: a $10\times 10\times
10 $ lattice with 200 configurations. (b). The difference between the DSHCs
of sizes $6\times 6\times 6$ and $10\times 10\times 10$. (c) and (d) are
similar to (a) and (b), for $W=3$ (the right side peak) and $W=10$ (the left
side peak), respectively. The insets in (a) and (c) shows the DSHCs at $%
10\times 10\times 10$ for $W=0.4$ and $10,$ respectively, which are fit by
the function $\frac{a}{|\protect\sigma _{SH}^{1}-\protect\sigma _{SH}|^{b}+c}
$. }
\label{fig:finite_size}
\end{figure}
Fig. \ref{fig:various_impurity} (a) shows the DSHC at $W=3$ for a $6\times
6\times 6$ lattice, with the Fermi energy $E_{f}$ fixed as the same value as
in Fig. 1. Here the open squares correspond to the result obtained over $%
N=1000$ configurations of random disorder and BCs, while the closed squares
are for the $N=5000$ configurations. Clearly the DSHC becomes smoother with
the increase of $N$, whose symmetric peak position remains unchanged with
essentially the same envelop. The solid curve in Fig. \ref%
{fig:various_impurity} (a) is obtained by averaging the DSHC at $N=1000$
over a small range of $\sigma $: $[\sigma -\delta ,\sigma +\delta ]$ with $%
\delta =0.2$ ($e/8\pi )$, defined by $P_{a}(\sigma )\equiv \frac{1}{2\delta }%
\int_{\sigma -\delta }^{\sigma +\delta }d\sigma ^{\prime }P(\sigma ^{\prime
})$, which coincides with the data at $N=5000$ very well and is plotted in a
wider range of $\sigma $ in the inset. The calculated $P_{a}$ at $W=0,$ $%
0.4, $ $3.0,$ and $10,$ respectively, for a $8\times 8\times 8$ lattice
averaged over 200 configurations are presented in Fig. \ref%
{fig:various_impurity} (b). The main panel focus on the neighborhood around
the peaks of $P_{a}(\sigma )$ and the inset illustrates the peaks in a
larger scale, whose lineshapes are generally symmetric such that one may
read off the value of $\sigma _{SH}$ directly from the peak position as
discussed above.

Now we study the sample-size dependence of the SHC with focusing on three
disorder strengths: $W=0.4$ for the weak disorder regime; $W=3$ for the
intermediate regime; and $W=10$ for the strong disorder regime. The results
for $W=0.4$ are plotted in Fig. \ref{fig:finite_size} (a) for three
different sizes of the lattice: $6\times 6\times 6$ with $N=500$ (the solid
curve), $8\times 8\times 8$ with $N=200$ (the dashed), and $10\times
10\times 10$ with $N=200$ (the dotted). In Fig. \ref{fig:finite_size} (b)
the difference between the $6\times 6\times 6$ and $10\times 10\times 10$
lattices is also presented. As the integrated $P_{a}$ is normalized to unit
for all sizes, the $P_{a}$ at $10\times 10\times 10$ has relatively much
longer tail such that $\Delta P_{a}$ in Fig. 3(b) remains negative over a
wide range of $\sigma _{SH}^{1}$ that is not easily seen by a naked eye.
These results show that the peak of $P_{a}(\sigma _{SH}^{1})$ is
significantly broadened and reduced with the increase of the sample size,
thus the fluctuations may survive in the large system size limit, like near
the critical point of quantum Hall system \cite{Sora_cho}. Such large
fluctuations of the SHC may be attributed to the nonconserved spins under
the random scattering of disorder. However, the peak position remains
essentially unchanged, which still well decides the averaged $\sigma _{SH}$.
On the other hand, much less sample-size dependence is found for $W=3$ and $%
W=10$, corresponding to two peaks in Fig. \ref{fig:finite_size} (c),
respectively, where the data for different sample-sizes all coincide with
each other. And the differences of the DSHCs between the $6\times 6\times 6$
and $10\times 10\times 10$ lattices are shown in Fig. \ref{fig:finite_size}
(d) which are indeed much reduced as compared the weak disorder case in (b).

\begin{figure}[tbph]
\resizebox{60mm}{!}{    \includegraphics{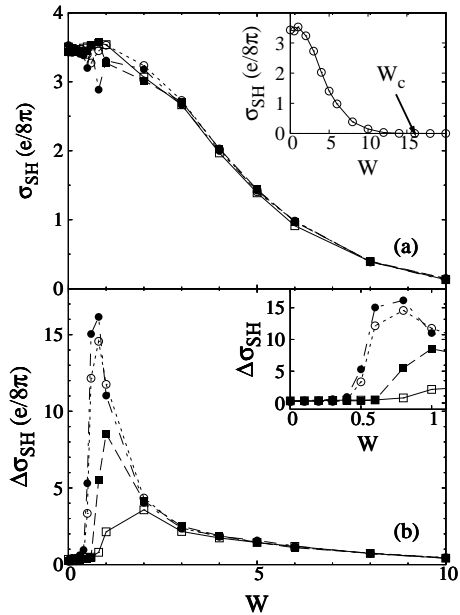} } \centering
\caption{(a) The $W$ dependence of $\protect\sigma _{SH}$. The solid curve
with open squares is for $6\times 6\times 6$, the dashed curve with close
squares is for $8\times 8\times 8$, and the dotted curve with open circles
is for $10\times 10\times 10$, while the dash-dot curve with closed circles
is for $6\times 6\times 24$. Inset: $\protect\sigma _{SH}$ over a wider
disorder regime, calculated for a $10\times 10\times 10$ lattice, where $%
W_{c}$ denotes the critical disorder for 3D Anderson localization (see
text). (b) The $W$ dependence of the HWHM $\Delta \protect\sigma _{SH}$. The
inset shows the details in a weak disorder regime. The notations are the
same as in the main panel of (a). }
\label{fig:shc_vs_impurity}
\end{figure}
To further characterize the size-dependence of $P_{a}$ and $\sigma _{SH}$,
we use a function $\frac{a}{|\sigma _{SH}^{1}-\sigma _{SH}|^{b}+c}$ to fit $%
P_{a}$, in which the typical width of the DSHC, $\Delta \sigma_{SH} $,
defined as the half width half maximum (HWHM), is given by $c^{1/b}$, and
two examples of the good fitting are shown in the insets of Fig. \ref%
{fig:finite_size} (a) and (c) for a\ $10\times 10\times 10$ lattice. In this
way we can systematically determine both $\sigma _{SH}$ and the
corresponding $\Delta \sigma _{SH}$ at different sample sizes and disorder
strengths. The results are depicted in Figs. \ref{fig:shc_vs_impurity} (a)
and (b) as functions of the disorder strength $W$. In the weak disorder
regime, we see that $\sigma _{SH}$ remains almost the same as the pure
system. With the increase of $W$, $\sigma _{SH}$ decreases monotonically and
is reduced to $5\%$ of the disorder-free value at $\ W\sim 10$. It becomes
indistinguishable from zero around $W\sim 14$, which is quite close to the
typical critical disorder strength, $W_{c}\sim 16$ \cite{3D_Wc}, of the
Anderson localization in 3D systems as marked in the inset of Fig. \ref%
{fig:shc_vs_impurity} (a). So the results suggest that the SHE always occurs
in the delocalized regime below $W_{c}.$ Furthermore, the finite-size effect
is very weak at $W\geq 2,$ from $6\times 6\times 6$ to $10\times 10\times 10$%
, with the continuous reduction of $\Delta \sigma _{SH}$ [Fig. \ref%
{fig:shc_vs_impurity} (b)].

The overall fluctuations of the SHC and the sample-size dependence are the
strongest in the intermediate regime of $0.5<W<2$. Both effects are then
monotonically reduced at $W\geq 2$, where $\sigma _{SH}$ becomes weakly
dependent on the sample size and the relatively small $\Delta \sigma _{SH}$
also indicates the reduction of intrinsic fluctuations at larger $W$. These
results suggest that there may exist a characteristic length scale in the
spin transport which \emph{decreases} with the increase of $W$. If this is
true, then the extrapolation to the thermodynamic limit should be at least
reliable in the strong disorder regime for the present finite-size
calculation. By a simple interpolation between the pure case and the strong
disorder case, then the SHC is expected to be robust over a wide range of
disorder strength.

In conclusion, we numerically studied the distribution of the spin Hall
conductance and determine the SHC. The main result shown in Fig. \ref%
{fig:shc_vs_impurity} indicates that in the weak disorder regime $\sigma
_{SH}$ remains almost the same as the value for the pure system. With the
increase of the disorder strength, $\sigma _{SH}$ is reduced and terminates
before the 3D Anderson localization takes place. Although the calculation
has been performed on finite lattice sizes, through the finite-size analysis
of the distribution function of the SHC, we found that the results are quite
size-independent, suggesting that the SHC in the 3D Luttinger model be
robust.  This is in contrast to the
vanishing behavior found in 2D electron Rashba model,
in agreement with analytical results considering vertex
corrections\cite{vertex2,vertex_luttinger}.

\begin{acknowledgments}
We thank the helpful discussion with S.C. Zhang, M.W. Wu, and C.X. Liu. \
This work is partially supported by the grants from the NSFC, ACS-PRF
41752-AC10, and the NSF grant/DMR-0307170. \ The computation of this project
was performed on the HP-SC45 Sigma-X parallel computer of ITP and ICTS, CAS.
\end{acknowledgments}

\end{document}